\documentclass[a4paper,12pt]{article}
\usepackage{amsmath, amsthm, amsfonts, amssymb, graphics}
\usepackage[mathscr]{eucal}

\begin{document}

%
%

\begin{center}{\bf R.R. Gadyl'shin\footnote{Bashkir State
Pedagogical University, Ufa, Russia. E-mail: gadylshin@bspu.ru}}
\end{center}
\medskip

\begin{center}
{\bf ON LOCAL PERTURBATIONS OF SCHR\"ODINGER OPERATOR ON PLANE}
\end{center}
\medskip

\begin{quote}\quad
{\small We obtain necessary and sufficient conditions for
emerging of small eigenvalue for Schr\"odinger operator on plane
under local operator perturbations. In the case the eigenvalue
emerges we construct its asymptotics. The examples are given.

}
\end{quote}
\medskip

\centerline{\bf 1.~INTRODUCTION}
\medskip

The questions of existence of bound states and asymptotics for
associated eigenvalues (if they exist) for Schr\"odinger
operator are investigated in [1]--[10]. In [11] in order to
study the case of small perturbation of Schr\"odinger operator
in the axis originated by sufficient arbitrary localized
second-order operator it was used an approach differing from
ones employed in [1]--[10]. This approach gave a simple
explanation of "non-regular" (non-obligatory) emerging of
eigenvalues under, obviously, regular perturbation; the essence
of this approach is as follows. Instead of spectral parameter
$\lambda$ they introduced more natural frequency one $k$ related
with spectral parameter by equality $\lambda=-k^2$, where $k$
belonged to complex half-plane $\hbox{\rm Re}\, k>0$. The
solutions of both non-perturbed and perturbed equation were
continued into complex plane by frequency parameter. Under such
continuation the solution of non-perturbed equation had a pole
at zero moving under perturbation and the residue at this pole
(both for perturbed and non-perturbed problems) was a solution
of corresponding homogeneous equation. For non-perturbed problem
this residue is a constant considered as exponent in a power
$-kt$ with $k=0$. The side the pole moved to determined the
exponential increasing or decreasing of residue for perturbed
problem. As a result, if pole moved into half-plane
$\mathbb{C}_+\overset{def}{=}\{k:\,\hbox{\rm Re}\, k>0\}$, the
eigenvalue had emerged, and it had not in the case of shift into
the half-plane $\hbox{\rm Re}\,\, k\le0$. Direction of this
shift depended on the operator of perturbation.

For the small perturbation of Schr\"odinger operator on plane
considered in this work and carried out by localized
second-order operator, the situation changes noticeably, because
here the solution of non-perturbed equation has a logarithmic
branch point instead of pole at $k=0$, and, therefore, is
analytically continued from half-plane into complex plane with a
cut alone negative real semiaxis. None the less, following the
scheme proposed in [11], it is possible to get the necessary and
sufficient conditions under those the solution of perturbed
equation has a small pole in a right half-plane (or, by other
words, a small eigenvalue emerge) and to calcualte the
asymptotics for it.

The construction of proof, on the one hand, being close to [11],
and, on the other hand, having its some differences (regarding
to the "replacement" of pole to logarithmic singularity at
non-perturbed equation), it is convenient to keep where possible
the notations and style of exposition of work [11].

The structure of the work is the following one. In the second
section we give a formulation of the main statement (Theorem 1)
and its corollaries. The third section is devoted to the proof
of the main statement. In concluding fourth section we show some
examples illustrating the main statement.

\medskip

\begin{center} {\bf 2.~FORMUALTION OF THE MAIN RESULT}
\end{center}

\medskip

Hereafter, $W_{2,loc}^j({\mathbb{R}^2})$ is a set of functions
defined on ${\mathbb{R}^2}$ whose restrictions into each bounded
domain $D\subset {\mathbb{R}^2}$ belong to $W_2^j(D)$,
$\left\|\bullet\right\|_G$ and $\left\|\bullet\right\|_{j,G}$
are norms in $L_2(G)$ and $W_2^j(G)$, respectively. Next, let
$Q$ be an arbitrary bounded domain in  ${\mathbb{R}^2}$,
$L_2({\mathbb{R}^2};Q)$ be a subset of functions from
$L_2({\mathbb{R}^2})$ having support in $\overline {Q}$,
${\mathcal L}_\varepsilon$ be a linear operator mapping
$W_{2,loc}^j({\mathbb{R}^2})$ into $L_2({\mathbb{R}^2};Q)$ such
that $\left\|{\mathcal L}_\varepsilon[u]\right\|_{Q}\le
C({\mathcal L})\,\left\|u\right\|_{2,Q}$, where constant
$C({\mathcal L})$ is independent on $\varepsilon$,
$0<\varepsilon\ll1$,
$$
\begin{array}{l}
\left<g\right>=\int\limits_{\mathbb{R}^2} g\,dx,\qquad
H_0=-\Delta,\qquad
H_\varepsilon=-\left(\Delta+\varepsilon{\mathcal
L}_\varepsilon\right).
\end{array}
$$
Hereinafter,  $x=(x_1,x_2)$, $y=(y_1,y_2)$. We define linear
operators $A(k)\,:\,L_2({\mathbb{R}^2};Q)\to
W^2_{2,loc}({\mathbb{R}^2})$ and
$T_\varepsilon^{(0)}(k)\,:\,L_2({\mathbb{R}^2};Q)\to
L_2({\mathbb{R}^2};Q)$ in a following way:
$$
\begin{aligned} A(k)g&=-\frac{1}{2\pi}\int\limits_{{\mathbb R}^2} {\mathcal
K}_0(k|x-y|)g(y)\,dy,\\ T_\varepsilon^{(0)}(k)g&={\mathcal
L}_\varepsilon[A(k)g]-\frac{\left<g\right>{\mathcal
L}_\varepsilon[1]}{2\pi}\left(\ln k+{\bf C}-\ln 2\right),
\end{aligned}
$$
where ${\mathcal K}_0$ is a Basse function of zero order (i.e.,
${\mathcal K}_0(z)=\frac{i\pi}{2} {\mathcal H}_0^{(1)}(iz)$,
where ${\mathcal H}^{(1)}_0$ is the Hankel function of first
kind and zero order), ${\bf C}$ is the Euler constant. We denote
by $\mathcal{B}(X,Y)$ the Banach space of linear bounded
operators from Banach space $X$ into Banach space $Y$, $\mathcal
{B}(X)\overset{def}{=}\mathcal{B}(X,X)$, and we will employ the
symbol $\mathcal{B}^h(X,Y)$ ($\mathcal{B}^h(X)$) for the set of
holomorphic operator-valued functions with values in
$\mathcal{B}(X,Y)$ (in $\mathcal{B}(X)$). Let $I$ be the
identity operator and $S(t)$ be a circle of radius $t$ with
center at zero in $\mathbb{C}$ and let
$S_\pm(t)=S(t)\backslash\mathbb{R}^\pm$. Hereinafter $\mathbb
{R}^-$ ($\mathbb {R}^+$) is a non-positive (non-negative) real
semiaxis in a complex plane. Since
$$
{\mathcal K}_0(z)=-\ln
\frac{z}{2}\,\sum\limits_{j=0}^\infty\frac{z^{2j}}{4^j(j!)^2}+
\sum\limits_{j=0}^\infty\frac{z^{2j}\psi(j+1)}{4^j(j!)^2},
$$
where $\psi(z)$ is a logarithmic derivation of Gamma function
(see, for instance, [12], [13]), the definition of $A(k)$ and
$T_\varepsilon^{(0)}(k)$ immediately implies

{\bf Lemma 1.} {\it For each $R>0$ there exists
$\varepsilon_0(R)>0$ such that for each fixed
$\varepsilon<\varepsilon_0(R)$

1) there exists $B_\varepsilon(k)\overset{def}{=}(I+\varepsilon
T_\varepsilon^{(0)}(k))^{-1}\in
\mathcal{B}^h(L_2(\mathbb{R}^2;Q))$ as $k\in
 S_-(R)$, and also, the uniform convergence
$B_\varepsilon(k)\underset{\varepsilon\to0}{\to}I$ takes place;

2) there exists at most one  (for each fixed $\varepsilon$)
solution $k_\varepsilon\in S_-(R)$ of equation
$$
1+\varepsilon\frac{1}{2\pi}\left<B_\varepsilon(k){\mathcal
L}_\varepsilon[1]\right>\left(\ln k+{\bf C}-\ln 2\right)=
0,\eqno(2.1)
$$
moreover, if it exists, then
$$
k_{\varepsilon}\underset{\varepsilon\to0}{\to}0.\eqno{(2.2)}
$$
}

{\bf Remark 1.} The statement of item  2)  does not exclude the
situation, when for some values $\varepsilon$ the solution
$k_\varepsilon\in S_-(R)$  of equation (2.1) exists, and for
others it doesn't.

We will say the operator $\mathcal{L}_\varepsilon$ to be
real-valued if $\hbox{\rm
Im}\,\left<\overline{g}\mathcal{L}_\varepsilon[g]\right>=0$ for
each function $g\in W_{2,loc}^2(\mathbb{R}^2)$. Let us denote
$\Pi_s(t)=\{k:\,|\hbox{\rm Im}\,\,k|< sC({\mathcal
L}),\,\,\hbox{\rm Re}\,\,k>t\}$. The main contents of this work
is the following statement whose proof next section is devoted
to.

{\bf Theorem 1.} {\it For $\varepsilon\to0$ there exists
$t(\varepsilon)\to\infty$, such that in
$\mathbb{C}\backslash\Pi_\varepsilon(t(\varepsilon))$ for each
fixed $\varepsilon$ there is at most one eigenvalue
$\lambda_\varepsilon$ of operator $H_\varepsilon$. The criteria
of existence of this eigenvalue is an existence of solution
$k_\varepsilon\in \mathbb{C}_+$ for the equation (2.1),
moreover,
$$
\lambda_\varepsilon=-k_\varepsilon^2,\eqno(2.3)
$$
and associated unique eigenfunction $\phi_\varepsilon$ is of the
form
$$
\phi_\varepsilon=A(k_\varepsilon)B_\varepsilon(k_\varepsilon){\mathcal
L}_\varepsilon[1].\quad \eqno(2.4)
$$

If, in addition, the operator ${\mathcal L}_\varepsilon$ is
real, then the statement of the theorem is true outside the
interval $(t(\varepsilon),\infty)$ of real axis.}

The assertions (2.1), (2.2) and statement of Theorem 1 yield

{\bf Corollary.} {\it If $\mathcal{L}_\varepsilon[1]=0$, then
there is no $\lambda_\varepsilon\notin
\mathbb{C}\backslash\Pi_\varepsilon(t(\varepsilon))$.

If there is no  $\lambda_\varepsilon\notin
\mathbb{C}\backslash\Pi_\varepsilon(t(\varepsilon))$, then
$\lambda_\varepsilon\to0$ as
$\varepsilon\to0$.}

On the one hand, Theorem 1 give neither necessary nor sufficient
conditions for operator $\mathcal{L}_\varepsilon$ under those
perturbed operator $H_\varepsilon$ has small eigenvalues (except
necessary condition given in Corollary), and, moreover, this
theorem provides no asymptotics for these eigenvalues if they
emerge. On the other hand, solving equation (2.1), one can
easily deduce these conditions and asymptotics. Indeed, bearing
in mind the definition of operators $B_\varepsilon(k)$,
$T_\varepsilon^{(0)}(k)$, $A(k)$ and
$$
\left(\Delta^{-1}g\right)(x)\overset{def}{=}\frac{1}{2\pi}\int\limits_{{\mathbb
R}^2}\ln |x-y|\,g(y)\,dy,
$$
one obtain the correctness of the following statement.

{\bf Lemma 2.} {\it If there exists a solution $k_\varepsilon\in
 S_-(R)$ of equation (2.1), then
$$
k_\varepsilon=\exp\left(-M(\varepsilon)\right), \eqno(2.5)
$$
where $M(\varepsilon)$ has an asymptotics determined by equality
$$
M{(\varepsilon)}=\frac{2\pi}{\varepsilon\sum\limits_{j=0}^\infty
\left(-\varepsilon\right)^j\left<\left({\mathcal
L}_\varepsilon\circ\Delta^{-1}\right)^j{\mathcal
L}_\varepsilon[1]\right>}+{\bf C}-\ln 2 \eqno(2.6)
$$
as $\varepsilon\to0$, and this asymptotics holds on power
precision.

If for the function $\widetilde{M}(\varepsilon)$ equalling to
right hand of (2.6) there exists  $\alpha>0$ such that for
$\varepsilon\to0$  the inequalities
$$
\hbox{\rm
Re}\,\widetilde{M}(\varepsilon)>\alpha\varepsilon^{\alpha}, \qquad
 \left|\hbox{\rm Im}\,\widetilde{M}(\varepsilon)\right|
<\frac{\pi}{2}-\alpha\varepsilon^{\alpha},\eqno(2.7)
$$
hold, then for each $R>0$ there exists $\varepsilon_0(R)>0$ such
that for  $\varepsilon<\varepsilon_0(R)$ there exist solutions
$k_{\varepsilon}\in S^+(R)\overset{def}{=} S(R)\cap
\mathbb{C}_+$ of equation (2.1).

If for the function $\widetilde{M}(\varepsilon)$ equalling to
right hand of (2.6) there exists $\alpha>0$ such that for
$\varepsilon\to0$  the inequalities
$$
\hbox{\rm
Re}\,\widetilde{M}(\varepsilon)<-\alpha\varepsilon^{\alpha}\eqno(2.8)
$$
or
$$
\left|\hbox{\rm Im}\,\widetilde{M}(\varepsilon)\right|
>\frac{\pi}{2}+\alpha\varepsilon^{\alpha},\eqno(2.9)
$$
hold, then for each $R>0$ there exists $\varepsilon_0(R)>0$ such
that for  $\varepsilon<\varepsilon_0(R)$ there is no solutions
$k_{\varepsilon}\in S^+(R)$ of equation (2.1).}

In its turn, from Lemma 2 and Theorem 1 it follows

{\bf Theorem 2.} {\it If for function
$\widetilde{M}(\varepsilon)$ equalling to right hand (2.6) there
exists a number $\alpha\ge\beta>0$ such that the inequalities
(2.7)  are true, then there exists $t(\varepsilon)\to\infty$
such that in
$\mathbb{C}\backslash\Pi_{\varepsilon}(t(\varepsilon))$ there is
one eigenvalue $\lambda_{\varepsilon}$ of operator
$H_{\varepsilon}$ and it has asymptotics given by (2.3), (2.5),
(2.6).

If for the function $\widetilde{M}(\varepsilon)$ equalling to
right hand of (2.6) there exists a number $\alpha>0$ such that
one of inequalities  (2.8), (2.9) holds then there exists
$t(\varepsilon)\to\infty$, such that in
$\mathbb{C}\backslash\Pi_{\varepsilon}(t(\varepsilon))$ there is
no eigenvalue of operator $H_{\varepsilon}$.}

In particular, it follows from Theorem 2, that if the eigenvalue
exists, it has asymptotics
$$
\lambda_\varepsilon=-4\exp\left\{-\frac{4\pi}{\varepsilon\sum\limits_{j=0}^\infty
\left(-\varepsilon\right)^j\left<\left({\mathcal
L}_\varepsilon\circ\Delta^{-1}\right)^j{\mathcal
L}_\varepsilon[1]\right>}-2{\bf C}\right\}.
$$

{\bf Remark 2.} The formulation of Theorem 2 means that if one
of its assumption takes place only on some subsequence
$\varepsilon_j\to0$, then the corresponding statement holds true
on this subsequence.

\medskip
\begin{center}
{\bf 3.~PROOF OF THEOREM 1}
\end{center}

\medskip

We indicate by $\mathcal{B}^m(X,Y)$ (by $\mathcal{B}^m(X)$) the
set of meromorphic operator-valued functions with values in
$\mathcal{B}(X,Y)$ (in $\mathcal{B}(X)$). Let
$\mathcal{B}(X,W_{2,loc}^2(\mathbb{R}^2))$ be a set of linear
operators from Banach space $X$ in $W_{2,loc}^2(\mathbb{R}^2)$
whose restriction on each bounded subset $D$ belongs to
$\mathcal{B}(X,W_{2}^2(D))$. Similarly, by
$\mathcal{B}^h(X,W_{2,loc}^2(\mathbb{R}^2))$ (by
$\mathcal{B}^m(X,W_{2,loc}^2(\mathbb{R}^2))$) we denote the set
of operator-valued functions with values in
$\mathcal{B}(X,W_{2,loc}^2(\mathbb{R}^2))$ such that for each
bounded  $D$ they belong to $\mathcal{B}^h(X,W_{2}^2(D))$
(belong to $\mathcal{B}^m(X,W_{2}^2(D))$). Let
$P_\varepsilon(k)$ be an operator defined by equality
$$
P_\varepsilon(k)f=-\varepsilon \frac{\left(\ln k+{\bf C}-\ln
2\right)\left<B_\varepsilon(k)f\right>B_\varepsilon(k){\mathcal
L}_\varepsilon[1]} {2\pi+\varepsilon{\left(\ln k+{\bf C}-\ln
2\right) }\left<B_\varepsilon(k){\mathcal L}_\varepsilon[1]\right>
}+ B_\varepsilon(k)f,
$$
$$\mathcal{R}_\varepsilon(k)\overset{def}{=}A(k)P_\varepsilon(k).$$

{\bf Theorem 3.} {\it For each  $R>0$ there exists
$\varepsilon_0(R)>0$ such that for $\varepsilon<\varepsilon_0$
and  $k\in S_-(R)$

1) $\mathcal{R}_\varepsilon(k)\in
\mathcal{B}^m(L_2(\mathbb{R}^2;Q),W_{2,loc}^2(\mathbb{R}^2))$,
and also, poles $k_\varepsilon$ of operator
$\mathcal{R}_\varepsilon(k)$ coincide with solutions of equation
(2.1); if, in  addition,  $k\in \mathbb{C}_+$, then
$\mathcal{R}_\varepsilon(k)\in
\mathcal{B}^m(L_2(\mathbb{R}^2,Q);W_{2}^2(\mathbb{R}^2))$;

2) for each $f\in L_2(\mathbb{R}^2;Q)$ the function
$u_\varepsilon=\mathcal{R}_\varepsilon(k)f$ is a solution of the
equation
$$
-H_\varepsilon u_\varepsilon= k^2u_\varepsilon+f\quad\hbox{in
${\mathbb{R}^2}$};\eqno(3.1)
$$

3) the residue of function $u_\varepsilon$  at pole
$k_\varepsilon$ is determined by equality (2.4) up to
multiplicative constant, moreover, this constant is nonzero if
$\left<f\right>\not=0$.}

{\bf Proof.} By definition  $A(k)\in
\mathcal{B}^h(L_2(\mathbb{R}^2;Q),W_{2,loc}^2(\mathbb{R}^2))$ as
$k\in S_-(R)$ for each  $R$, moreover, $A(k)\in
\mathcal{B}^h(L_2(\mathbb{R}^2;Q),W_{2}^2(\mathbb{R}^2))$ as
 $k\in
\mathbb{C}_+$. Hence, taking into account the definition of
$\mathcal{R}_\varepsilon(k)$ and Lemma 1, we get the validity of
statement 1) of the theorem being proved.

Let us show the correctness of item 2). We seek the solution of
equation (3.1) as
$$
u_\varepsilon=A(k)g_\varepsilon,\eqno(3.2)
$$
where $g_\varepsilon$ is a function to be found belonging to
$L_2({\mathbb{R}^2};Q)$. Substituting  (3.2) into (3.1) and
bearing in mind that (3.2) is a solution of the equation
$$
-H_0 u_\varepsilon= k^2u_\varepsilon+g_\varepsilon\quad\hbox{in
${\mathbb{R}^2}$},
$$
we get that (3.2) is a solution of  (3.1) in the case when
$$
(I+\varepsilon T_\varepsilon(k))g_\varepsilon=f,\eqno(3.3)
$$
where
$$
T_\varepsilon(k)={\mathcal L}_\varepsilon A(k).\eqno(3.4)
$$
It follows from  (3.4) and definition of
$T^{(0)}_\varepsilon(k)$ and $A(k)$ that the operator
$T_\varepsilon(k)$ acts in a following way:
$$
T_\varepsilon(k)g=\frac{\left<g\right>{\mathcal
L}_\varepsilon[1]}{2\pi}\left(\ln k+{\bf C}-\ln
2\right)+T_\varepsilon^{(0)}(k)g. \eqno(3.5)
$$

Let $R>0$ be an arbitrary number and $\varepsilon$ satisfy to
Lemma 1. Applying the operator $B_\varepsilon(k)$ to both hands
of  (3.3) and taking into account (3.5), we obtain that
$$
\left(g_\varepsilon+\varepsilon
\frac{\left<g_\varepsilon\right>\left(\ln k+{\bf C}-\ln
2\right)} {2\pi}B_\varepsilon(k){\mathcal
L}_\varepsilon[1]\right)= B_\varepsilon(k)f.\eqno(3.6)
$$
Having integrated  (3.6), we get that
$$
\left<g_\varepsilon\right> \left(1+\varepsilon\frac{\left(\ln
k+{\bf C}-\ln 2\right) }{2\pi}\left<B_\varepsilon(k){\mathcal
L}_\varepsilon[1]\right>
\right)=\left<B_\varepsilon(k)f\right>.\eqno(3.7)
$$
Calculating $\left<g_\varepsilon\right>$ by (3.7) and
substituting this value in (3.6), we deduce that
$$
g_\varepsilon=P_\varepsilon(k)f.\eqno(3.8)
$$
Assertions  (3.2) and (3.8) imply the validity of statement 2).
In its turn, the correctness 3) is yielded by 1), 2) and
definition of $\mathcal {R}_\varepsilon(k)$. The proof is
complete.

We denote by $R_\varepsilon(\lambda)$ the resolvent for operator
$H_\varepsilon$.

{\bf Lemma 3.} {\it The number of poles of resolvent
$R_\varepsilon(\lambda)$, their orders and dimensions of residua
at them are completely determined by functions from
$L_2(\mathbb{R}^2;Q)$}.

{\bf Proof.} Let $F$ be an arbitrary function having finite
support $D$. It is easy to see that
$$
R_\varepsilon(\lambda)F=R_0(\lambda)F+\varepsilon
R_\varepsilon(\lambda) \mathcal{L}_\varepsilon R_0(\lambda)F.
\eqno (3.9)
$$
Since $R_0(\lambda)$ has no poles, and $\hbox{\rm
supp}\,\left(\mathcal{L}_\varepsilon R_0(\lambda)F\right)
\subset \overline {Q}$, it follows from (3.9) the statement of
lemma.

We will employ the symbol $\Sigma(H_\varepsilon)$ for the set of
eigenvalues of operator $H_\varepsilon$.

{\bf Theorem 4.} {\it Let $R>0$ be an arbitrary number and
$\varepsilon_0(R)$ satisfy to statement of Theorem 3,
$\lambda=-k^2$. Then
$$
\mathcal{R}_\varepsilon(k)f=-R_\varepsilon(\lambda)f\eqno(3.10)
$$
for each  $f\in L_2(\mathbb{R}^2;Q)$,
$0<\varepsilon<\varepsilon_0(R)$ and $k\in S^+(R)$ (or,
equivalently, for each $\lambda\in S_+(R^2)$).

If $k_\varepsilon\in  S_-(R)$ is a pole of operator
$\mathcal{R}_\varepsilon(k)$ and $\hbox{\rm Re}\,
k_\varepsilon\le0$, then $\Sigma(H_\varepsilon)\cap
S_+(R^2)=\emptyset$.

If $k_\varepsilon\in S^+(R)$ is a pole of operator
$\mathcal{R}_\varepsilon(k)$, then $\Sigma(H_\varepsilon)\cap
S_+(R^2)=\{\lambda_\varepsilon\}$, where $\lambda_\varepsilon$
and the associated unique eigenfunction are given by equalities
(2.3) and (2.4).}

{\bf Proof.}  The function $\lambda=-k^2$ mapping $S^+(R)$ onto
$S_+(R^2)$ in one-to-one manner, the correctness of equality
(3.10) follows from statement 2) of Theorem 3 and the definition
of resolvent. The correctness of other statement of the theorem
being proved follows from Theorem 3 and Lemma 3.

{\bf Lemma 4.} {\it
$\Sigma(H_\varepsilon)\subset\Pi_\varepsilon(-\varepsilon
C({\mathcal L}))$. If ${\mathcal L}_\varepsilon$ is a real
operator then $\Sigma(H_\varepsilon)\subset [-\varepsilon
C({\mathcal L}),\infty)$.}

{\bf Proof.} Let $\lambda_\varepsilon\in\Sigma(H_\varepsilon)$,
$\phi_\varepsilon$ be an associated eigenfunction normalized in
$L_2(\mathbb{R}^2)$. Therefore
$$
H_\varepsilon\phi_\varepsilon=\lambda_\varepsilon\phi_\varepsilon
\qquad\hbox{in $\mathbb {R}^2$}.\eqno(3.11)
$$
Since $H_\varepsilon U=H_0U$ outside $\overline {Q}$, it is easy
to show that  $\phi_\varepsilon\in W_2^2(\mathbb{R}^2)$.
Multiplying both hands of  (3.11) by
$\overline{\phi_\varepsilon}$ and integrating by parts we obtain
the equality
$$
\left\|\nabla\phi_\varepsilon\right\|^2_{\mathbb
{R}^2}-\varepsilon\left<\overline{\phi_\varepsilon}{\mathcal
L}_\varepsilon\phi_\varepsilon\right>=
\lambda_\varepsilon.\eqno(3.12)
$$
Taking imaginary and real part of  (3.12) and using an estimate
 $\left\|{\mathcal L}_\varepsilon\phi_\varepsilon\right\|_Q\le
C({\mathcal L})\left\|\phi_\varepsilon\right\|_{2,Q}$, we
conclude that the lemma holds true.

Let $[a,b]_{\mathbb R}$ be a segment $[a,b]$ of real axis.

{\bf Lemma 5.} {\it For each $R>0$ there exists
$\varepsilon_0(R)>0$ such that $[0,R]_{\mathbb R}\cap
\Sigma(H_\varepsilon)=\emptyset$ as
$\varepsilon<\varepsilon_0$.}

{\bf Proof.} We prove the lemma by reductio ad absurdum. It is
easy to see that without loss of generality it is sufficient to
consider the situation $[0,R]_{\mathbb R}\cap
\Sigma(H_{\varepsilon})=\lambda_\varepsilon\underset
{\varepsilon\to0}{\to} \lambda_*$ in the cases when all
$\lambda_\varepsilon\not=0$, and, conversely, when all
$\lambda_\varepsilon=0$. Clear, in the latter case
$\lambda_*=0$.

We start from first case. Since $H_\varepsilon U=H_0U$ outside
$\overline{Q}$, then for the associated eigenfunction
$\phi_\varepsilon$ the equality $\phi_\varepsilon=0$ holds as
$|x|=T$ for each $T$ sufficiently large (see., for instance,
[14, \S~3.10]). Hence, $\lambda_\varepsilon$ and
$\phi_\varepsilon$ are eigenelements of the boundary value
problem
$$
H_\varepsilon\phi_\varepsilon=
\lambda_\varepsilon\phi_\varepsilon\quad\hbox{as
$|x|<T$},\qquad\hbox{ $\phi_\varepsilon=0$\quad as
$|x|=T$}.\eqno(3.13)
$$
Since (3.13) is a regular perturbation of the limit problem
$$
H_0\phi_*= \lambda_*\phi_0\quad\hbox{as $|x|<T$},\qquad
\hbox{$\phi_*=0$\quad as $|x|=T$},\eqno(3.14)
$$
it follows that having chosen $T$ in such a way that $\lambda_*$
not to be an eigenvalue of (3.14), we get that
$\phi_\varepsilon$ can not satisfy to (3.13).

Let us proceed to the second case $\lambda_\varepsilon=0$. In an
obvious way from (3.11) and properties of ${\cal L_\varepsilon}$
it follows the estimate
$$
\begin{array}{l}
\left\|\Delta\phi_\varepsilon\right\|_{{\mathbb R}^2}\le
\varepsilon C\left\|\phi_\varepsilon\right\|_{2,Q},\\
\left\|\nabla \phi_\varepsilon\right\|^2_{{\mathbb R}^2}\le
\varepsilon
C\left\|\phi_\varepsilon\right\|_Q\left\|\phi_\varepsilon\right\|_{2,Q}.
\end{array}
\eqno(3.15)
$$
Without loss of generality we assume that the function
$\phi_\varepsilon$ is normalized in $W_2^2(Q)$. Then the
estimates (3.15) imply that on $Q$
$$
\phi_\varepsilon=|Q|^{-1/2}+\psi_\varepsilon,\eqno(3.16)
$$
$$
\left\|\psi_\varepsilon\right\|_{1,Q}\to0\quad \hbox{as
$\varepsilon\to0$}.
\eqno(3.17)
$$
By localness of ${\cal L}_\varepsilon$ and (3.16) outside $Q$
the function $\phi_\varepsilon$ being a solution of boundary
value problem
$$
\Delta\phi_\varepsilon=0\qquad\hbox{in ${\mathbb R}^2\backslash
\overline{Q}$}\qquad
\phi_\varepsilon=|Q|^{-1/2}+\psi_\varepsilon\quad\hbox{on
$\partial Q$},
$$
it follows from (3.17) that $\phi_\varepsilon$ does not tends to
zero at infinity. Therefore, $\phi_\varepsilon\notin
L_2(\mathbb{R}^2)$. The contradiction obtained completes the
proof of lemma.

One can easily see that Theorem 1 is a direct implication of
Theorem 4 and Lemmas  4 and 5.

\centerline{\bf 4. EXAMPLES}

{\bf Example 1.} Let $\mathcal{L}_\varepsilon[g]=Vg$, where
$V\in C^\infty_0(Q)$. It follows from (2.6) that for this case
$$
\widetilde{M}{(\varepsilon)}=\frac{2\pi} {\varepsilon
m(\varepsilon)}+{\bf C}-\ln 2, \eqno(4.1)
$$
where
$$
m(\varepsilon)=\left<V\right>+ \sum\limits_{j=1}^3
\left(-\varepsilon\right)^j
\left<\left(V\Delta^{-1}\right)^jV\right>+O(\varepsilon^4).
\eqno(4.2)
$$
We note, that by integrating by parts it is easy to show the
validity of equalities:
$$
\aligned &\left<U\Delta^{-1}U\right>=-\left<\left(\nabla\left(
\Delta^{-1}U\right)\right)^2\right>,\quad
\left<\left(U\Delta^{-1}\right)^2U\right>=\left<U\left(
\Delta^{-1}U\right)^2\right>,\\
&\left<\left(U\Delta^{-1}\right)^3U\right>=\left<\left(U
\Delta^{-1}U\right)\Delta^{-1} \left(U
\Delta^{-1}U\right)\right>,\quad \hbox{if
$\left<U\right>=0$}.\endaligned \eqno(4.3)
$$
The assertions (4.1)--(4.3) yield
$$
\widetilde{M}{(\varepsilon)}=\frac{2\pi}
{\varepsilon\left<V\right>}+\frac{2\pi\left<V\Delta^{-1}V\right>}
{ \left<V\right>^2} +{\bf C}-\ln 2+O(\varepsilon),\qquad\hbox{if
$\left<V\right>\not=0$},\eqno(4.4)
$$
$$
\aligned
\widetilde{M}{(\varepsilon)}=&\frac{2\pi}{\varepsilon^2\left<\left(\nabla\left(
\Delta^{-1}V\right)\right)^2\right>}-\frac{2\pi\left<V\left(
\Delta^{-1}V\right)^2\right>}{\varepsilon\left<\left(\nabla\left(
\Delta^{-1}V\right)\right)^2\right>^2}\\ &+
\frac{2\pi\left<\left(V \Delta^{-1}V\right)\Delta^{-1}\left(V
\Delta^{-1}V\right)\right>}{\left<\left(\nabla\left(
\Delta^{-1}V\right)\right)^2\right>^2}+\frac{2\pi\left<V\left(
\Delta^{-1}V\right)^2\right>^2} {\left<\left(\nabla\left(
\Delta^{-1}V\right)\right)^2\right>^3}\\&+{\bf C}-\ln
2+O(\varepsilon), \qquad \hbox{if $\left<V\right>=0$}.
\endaligned\eqno(4.5)
$$

First we consider classical situation $\hbox{\rm Im}\,V=0$. It
follows from (4.4) and Theorem 2 that if $\left<V\right><0$,
then there is no eigenvalue converging to zero, and if
$\left<V\right>\,>0$, then
$$
\lambda_\varepsilon= -\varkappa_\varepsilon^2
\exp\left\{-\frac{4\pi}{\varepsilon\left<V\right>}\right\},
\eqno(4.6)
$$
$$
\varkappa_\varepsilon=2\exp\left\{-
\frac{2\pi\left<V\Delta^{-1}V\right>}{ \left<V\right>^2}-{\bf
C}\right\}+O(\varepsilon).
$$
By analogy, from  (4.5) and Theorem 2 it follows that if
$\left<V\right>=0$, then the eigenvalue exists  and has the
asymptotics
$$
\lambda_\varepsilon= -\varkappa_\varepsilon^2
\exp\left\{-\frac{4\pi}{\varepsilon^2\left\|\nabla\left(\Delta
^{-1}V\right)\right\|_{\mathbb{R}^2}^2}+ \frac{4\pi\left<V\left(
\Delta^{-1}V\right)^2\right>}{\varepsilon\left\|\nabla\left(\Delta
^{-1}V\right)\right\|_{\mathbb{R}^2}^4}\right\},\eqno(4.7)
$$
$$
\varkappa_\varepsilon=2\exp\left\{ -\frac{2\pi\left<\left(V
\Delta^{-1}V\right)\Delta^{-1}\left(V
\Delta^{-1}V\right)\right>}{ \left\|\nabla\left(\Delta
^{-1}V\right)\right\|_{\mathbb{R}^2}^4}- \frac{2\pi\left<V\left(
\Delta^{-1}V\right)^2\right>^2} {\left\|\nabla\left(\Delta
^{-1}V\right)\right\|_{\mathbb{R}^2}^6}-{\bf
C}\right\}+O(\varepsilon).
$$
It should be noted that formulas (4.6), (4.7) are well-known
(see [1], [2]).

Now let us consider complex potential $V$. It follows from
Theorem 2 and (4.4) that if  $\hbox{\rm
Im}\,\left<V\right>\not=0$, then for all sufficiently small
$\varepsilon$ there is no eigenvalue outside
$\mathbb{C}\backslash\Pi_{\varepsilon}(t(\varepsilon))$ (i.e.,
there is no small eigenvalue). Observe, this situation differs
noticeably from one-dimensional case in [11], where the
sufficient condition for the existence of the eigenvalue
converging to zero  was  $\hbox{\rm Re}\left<V\right>\,>0$.

Let us show that for complex potential even the condition
$\left<V\right>\,>0$ is insufficient for existence of
eigenvalue. Let $V=v+ia\Delta v$, where $v\in C^\infty_0(Q)$ is
a real function,  $\left<v\right>>0$, and $a$ is an arbitrary
real constant unknown yet. It is easy to see that in this case
$$
\left<V\right>=\left<v\right>\,>0,\qquad
\left<V\Delta^{-1}V\right>=(1-
a^2)\left<v\Delta^{-1}v\right>+i2a\left\|v\right\|_{\mathbb{R}^2}^2.
$$
Then it follows from (4.4) that
$$
\aligned \hbox{\rm
Re}\,\widetilde{M}{(\varepsilon)}&=\frac{2\pi}{\varepsilon
\left<v\right>}+ \frac{2\pi(1- a^2)\left<v\Delta^{-1}v\right>}{
\left<v\right>^2}+{\bf C}-\ln 2+O(\varepsilon),\\ \hbox{\rm
Im}\,\widetilde{M}{(\varepsilon)}&=\frac{4\pi
a\left\|v\right\|_{\mathbb{R}^2}^2}{
\left<v\right>^2}+O(\varepsilon).\endaligned \eqno(4.8)
$$
It follows from (4.8) and Theorem 2  that for
$$
|a|>\frac{\left<v\right>^2}{8 \left\|v\right\|^2}
$$
there is no small eigenvalue, and for
$$
|a|<\frac{\left<v\right>^2}{8 \left\|v\right\|^2}
$$
such eigenvalue exists and has asymptotics (4.6), where
$$\aligned
\varkappa_\varepsilon=&2\exp\left\{\frac{2\pi(a^2-1)\left<v\Delta^{-1}v\right>}{
\left<v\right>^2}-{\bf C}\right\}\\
&\times\left(\cos\left(\frac{4a\pi\left\|v\right\|^2}{\left<v\right>^2}
\right)-i\sin\left(\frac{4a\pi\left\|v\right\|^2}{\left<v\right>^2}
\right)\right)+O(\varepsilon).
\endaligned\eqno(4.9)
$$

In particular, it follows from (4.6) and (4.9) that for
$$
\frac{\left<v\right>^2}{16
\left\|v\right\|^2}<|a|<\frac{\left<v\right>^2}{8
\left\|v\right\|^2}
$$
the eigenvalue lies in the right complex half-plane (in contrast
to the case of real potential, for that eigenvalue can lie only
in the negative real semiaxis).

{\bf Example 2.} Let $\mathcal{L}_\varepsilon[g]=V_\varepsilon
g$, where $V_\varepsilon=V+\varepsilon V_1$, and  $V,\,V_1$ are
real function having supports in  $Q$. Then by (2.6), (4.3) and
by  equalities  checked easily
$$
\aligned
&\left<V_1\Delta^{-1}V\right>=\left<V\Delta^{-1}V_1\right>,\qquad
\left<\left(V\Delta^{-1}\right)^2V_1\right>=
\left<V\left(\Delta^{-1}V\right)\left(\Delta^{-1}V_1\right)\right>,\\
& \left<V\Delta^{-1}\left(V_1\Delta^{-1}V\right)\right>=
\left<V_1\left(\Delta^{-1}V\right)^2\right>\qquad\hbox{as
$\left<V\right>=0$}
\endaligned
$$
$\widetilde{M}(\varepsilon)$ has the form (4.1), where
$$
m(\varepsilon)=\left<V\right>+\varepsilon
\left(\left<V_1\right>-\left<V\Delta^{-1}V\right>\right)
+O(\varepsilon^2),\quad\hbox{if $\left<V\right>\not=0$},
\eqno(4.10)
$$
and
$$
\aligned
\frac{m(\varepsilon)}{\varepsilon}=&\left\|\nabla\left(\Delta
^{-1}V \right)\right\|_{\mathbb{R}^2}^2+\left<V_1\right>-
\varepsilon\left\{2\left<V_1\Delta^{-1}V\right>-
\left<V\left(\Delta^{-1}V\right)^2\right>\right\}
\\&-
\varepsilon^2m_2+O(\varepsilon^3),\quad\hbox{if
$\left<V\right>=0$},
\endaligned\eqno(4.11)
$$
where
$$ \aligned m_2=&\left<V_1\Delta^{-1}V_1\right>+
\left<\left(V \Delta^{-1}V\right)\Delta^{-1} \left(V
\Delta^{-1}V\right)\right>
\\&-
\left<V\left(\Delta^{-1}V\right)\left(\Delta^{-1}V_1\right)\right>
-
\left<V_1\left(\Delta^{-1}V\right)^2\right>
-
\left<V_1\Delta^{-1}\left(V\Delta^{-1}V\right)\right>.
\endaligned\eqno(4.12)
$$
It follows from (4.1), (4.10), (4.11) that
$$
\widetilde{M}{(\varepsilon)}=\frac{2\pi}{\varepsilon
\left<V\right>}+
\frac{2\pi\left(\left<V\Delta^{-1}V\right>-\left<V_1\right>\right)}{
\left<V\right>^2}+{\bf C}-\ln 2+O(\varepsilon),\quad\hbox{if
$\left<V\right>\not=0$},\eqno(4.13)
$$
and
$$
\aligned \widetilde{M}{(\varepsilon)}=&\frac{2\pi}
{\varepsilon^2\left(\left\|\nabla\left(\Delta ^{-1}V
\right)\right\|_{\mathbb{R}^2}^2+\left<V_1\right>\right)} +
\frac{2\pi\left(2\left<V_1\Delta^{-1}V\right>-
\left<V\left(\Delta^{-1}V\right)^2\right>\right)}{\varepsilon\left(\left\|\nabla\left(\Delta
^{-1}V \right)\right\|_{\mathbb{R}^2}^2+\left<V_1\right>\right)^2}
\\ &+
\frac{2\pi\left(2\left<V_1\Delta^{-1}V\right>-
\left<V\left(\Delta^{-1}V\right)^2\right>\right)^2}{\left(\left\|\nabla\left(\Delta
^{-1}V \right)\right\|_{\mathbb{R}^2}^2+\left<V_1\right>\right)^3}
+\frac{2\pi m_2}{\left(\left\|\nabla\left(\Delta ^{-1}V
\right)\right\|_{\mathbb{R}^2}^2+\left<V_1\right>\right)^2}
\\&+{\bf
C}-\ln 2+O(\varepsilon),\quad\hbox{if $\left<V\right>=0$}.
\endaligned\eqno(4.14)
$$

It follows from  (4.13), (4.14) and Theorem 2 that the condition
$\left<V_\varepsilon\right>\ge0$ is sufficient for existence of
eigenvalue. If $\left<V_\varepsilon\right>\,>0$ ''in principle''
(i.e., $\left<V\right>\,>0$), then by  (4.13) it has asymptotics
(4.6), where
$$
\varkappa_\varepsilon=2\exp
\left\{-\frac{2\pi\left(\left<V\Delta^{-1}V\right>-\left<V_1\right>\right)}
{\left<V\right>^2}-{\bf C}\right\}+O(\varepsilon).
$$
If $\left<V_\varepsilon\right>\,>0$ in small (i.e.,
$\left<V\right>\,=0$, but $\left<V_1\right>\,>0$), then due to
(4.14) the eigenvalue has asymptotics
$$
\aligned \lambda_\varepsilon=&-\varkappa_\varepsilon^2
\exp\Bigg\{-\frac{4\pi}{\varepsilon^2\left(\left<V_1\right>+
\left\|\nabla\left(\Delta^{-1}
V\right)\right\|_{\mathbb{R}^2}^2\right)}\\&\qquad\qquad\quad-
\frac{4\pi\left(2\left<V_1\Delta^{-1}V\right>-
\left<V\left(\Delta^{-1}V\right)^2\right>\right)}
{\varepsilon\left(\left\|\nabla\left(\Delta ^{-1}V
\right)\right\|_{\mathbb{R}^2}^2+\left<V_1\right>\right)^2}\Bigg\},
\endaligned
$$
$$
\aligned \varkappa_\varepsilon=2\exp
\Bigg\{&-\frac{2\pi\left(2\left<V_1\Delta^{-1}V\right>-
\left<V\left(\Delta^{-1}V\right)^2\right>\right)^2}
{\left(\left\|\nabla\left(\Delta ^{-1}V
\right)\right\|_{\mathbb{R}^2}^2+\left<V_1\right>\right)^3}\\&-
\frac{2\pi m_2}{\left(\left\|\nabla\left(\Delta ^{-1}V
\right)\right\|_{\mathbb{R}^2}^2+\left<V_1\right>\right)^2}-{\bf
C}\Bigg\}+O(\varepsilon),
\endaligned
$$
where constant $m_2$ is defined by (4.12). And, finally, if
$\left<V_\varepsilon\right>=0$, then in view of (4.14) and (4.3)
the eigenvalue has asymptotics
$$
 \lambda_\varepsilon=-\varkappa_\varepsilon^2
\exp\Bigg\{-\frac{4\pi}{\varepsilon^2
\left\|\nabla\left(\Delta^{-1} V\right)\right\|_{\mathbb{R}^2}^2}-
\frac{4\pi\left(2\left<V_1\Delta^{-1}V\right>+
\left<V\left(\Delta^{-1}V\right)^2\right>\right)}
{\varepsilon\left\|\nabla\left(\Delta ^{-1}V
\right)\right\|_{\mathbb{R}^2}^4}\Bigg\},
$$
$$
\varkappa_\varepsilon=2\exp
\Bigg\{-\frac{2\pi\left(2\left<V_1\Delta^{-1}V\right>+
\left<V\left(\Delta^{-1}V\right)^2\right>\right)^2}
{\left\|\nabla\left(\Delta ^{-1}V
\right)\right\|_{\mathbb{R}^2}^6}- \frac{2\pi
m^{(2)}}{\left\|\nabla\left(\Delta ^{-1}V
\right)\right\|_{\mathbb{R}^2}^4}-{\bf C}\Bigg\} +O(\varepsilon),
$$
where
$$
\aligned m^{(2)}=&-\left\|\nabla \left(
\Delta^{-1}V_1\right)\right\|^2+ \left<\left(V
\Delta^{-1}V\right)\Delta^{-1} \left(V \Delta^{-1}V\right)\right>
\\&-
\left<V\left(\Delta^{-1}V\right)\left(\Delta^{-1}V_1\right)\right>
-
\left<V_1\left(\Delta^{-1}V\right)^2\right>
-
\left<V_1\Delta^{-1}\left(V\Delta^{-1}V\right)\right>.
\endaligned
$$

However, in contrast to classic case (real $V_\varepsilon=V$)
the condition $\left<V_\varepsilon\right>\,<0$ is insufficient
for absence of eigenvalue. Indeed, let $v$ be a real function,
$V=\Delta v$, $V_1=av$, $\left<v\right>>0$ and $a$ is a some
arbitrary constant. Then by (4.14)
$$
\aligned \widetilde{M}{(\varepsilon)}=&\frac{2\pi}
{\varepsilon^2\left(\left\|\nabla v
\right\|_{\mathbb{R}^2}^2+a\left<v\right>\right)} +
\frac{2\pi\left(2a\left\|\nabla v \right\|_{\mathbb{R}^2}^2-
\left<v^2\Delta v\right>\right)}
{\varepsilon\left(\left\|\nabla\left(\Delta ^{-1}V
\right)\right\|_{\mathbb{R}^2}^2+\left<V_1\right>\right)^2}
\\ &+
\frac{2\pi\left(2a\left\|\nabla v \right\|_{\mathbb{R}^2}^2-
\left<v^2\Delta v\right>\right)^2}{\left(\left\|\nabla v
\right\|_{\mathbb{R}^2}^2+a\left<v\right>\right)^3} +\frac{2\pi
\widetilde{m}_2}{\left(\left\|\nabla v
\right\|_{\mathbb{R}^2}^2+a\left<v\right>\right)^2}
\\&+{\bf
C}-\ln 2+O(\varepsilon),
\endaligned\eqno(4.15)
$$
where
$$
\aligned \widetilde{m}_2=&a^2\left<v\Delta^{-1}v\right>+
\left<\left(v \Delta v\right)\Delta^{-1} \left(v \Delta v
\right)\right>\\& -a \left<v\left(\Delta^{-1}v\right)\Delta
v\right>
-
a\left<v^3\right>- a\left<v\Delta^{-1}\left(v\Delta v
\right)\right>.
\endaligned\eqno(4.16)
$$
It follows from (4.15) and Theorem 2 that for
$$
-\frac{\left\|\nabla v\right\|^2}{\left<v\right>}<a<0
$$
the eigenvalue exists and has asymptotics
$$
\lambda_\varepsilon=-\varkappa_\varepsilon^2
\exp\Bigg\{-\frac{4\pi}{\varepsilon^2 \left(\left\|\nabla v
\right\|_{\mathbb{R}^2}^2+a\left<v\right>\right)}-
\frac{4\pi\left(2a\left\|\nabla v \right\|_{\mathbb{R}^2}^2-
\left<v^2\Delta v\right>\right)} {\varepsilon\left(\left\|\nabla
v \right\|_{\mathbb{R}^2}^2+a\left<v\right>\right)^2}\Bigg\},
$$
$$
\varkappa_\varepsilon=2\exp
\Bigg\{-\frac{2\pi\left(2a\left\|\nabla v
\right\|_{\mathbb{R}^2}^2- \left<v^2\Delta v\right>\right)^2}
{\left(\left\|\nabla v
\right\|_{\mathbb{R}^2}^2+a\left<v\right>\right)^3}- \frac{2\pi
\widetilde{m}_2}{\left(\left\|\nabla v
\right\|_{\mathbb{R}^2}^2+a\left<v\right>\right)^2}-{\bf C}\Bigg\}
+O(\varepsilon),
$$
where $\widetilde{m}_2$ is defined in (4.16). At the same time,
clearly, $\left<V_\varepsilon\right>=a\left<v\right><0$ for the
indicated values of $a$. Similar effect was observed in [6].

{\bf Example 3.} Let
$$
\mathcal{L}_\varepsilon[g]=\chi(Q)\left<\rho_\varepsilon g\right>,
$$
where $\chi(Q)$ is a characteristic function of $Q$, (i.e., the
function equalling to one as $x\in Q$ and vanishing for other
$x$), and a function $\rho_\varepsilon$ be continuous in
$\overline{Q}$ and extended by zero outside $\overline{Q}$.

Since
$$
\mathcal{L}_\varepsilon[1]=\left<\rho_\varepsilon\right>,\eqno(4.17)
$$
it follows from Corollary of Theorem 1 that there is no small
eigenvalue if $\left<\rho_\varepsilon\right>=0$.

From definition of $\mathcal{L}_\varepsilon$ it follows that
$$
\left(\mathcal{L}_\varepsilon\circ\Delta^{-1}\right)^n
\mathcal{L}_\varepsilon[1]=\left<\rho_\varepsilon\Delta^{-1}
\chi\right>^n\chi,\qquad n\ge1.\eqno(4.18)
$$
Substituting (4.17), (4.18) into (2.6) we get that
$$
\widetilde{M}(\varepsilon)=\frac{2\pi}
{\varepsilon\left<\rho_\varepsilon\right>|Q|}
\left(1+\varepsilon\left<\rho_\varepsilon\Delta^{-1}\chi\right>\right)+{\bf
C}-\ln 2. \eqno(4.19)
$$

We remind, that in view of Theorem 2 the sufficient condition
for existence of small eigenvalue is an existence of $\alpha>0$
such that the inequalities (2.7) hold, due to  (4.19) they take
the form
$$
\aligned &\hbox{\rm
Re}\,\frac{1+\varepsilon\left<\rho_\varepsilon\Delta^{-1}\chi\right>}
{\varepsilon\left<\rho_\varepsilon\right>}
>\alpha\varepsilon^{\alpha}+\frac{|Q|\left(\ln 2-{\bf
C}\right)}{2\pi},\\&
 \left|\hbox{\rm Im}\,\frac{1+\varepsilon\left<\rho_\varepsilon
\Delta^{-1}\chi\right>} {\varepsilon\left<\rho_\varepsilon\right>}
\right|
<\frac{|Q|}{4}-\alpha\varepsilon^{\alpha}.\endaligned
\eqno(4.20)
$$
The first of these inequalities holds true, if quantities
$\left<\rho_\varepsilon \Delta^{-1}\chi\right>$ and
$\left<\rho_\varepsilon\right>$ are bounded by module as
$\varepsilon\to0$ and $\hbox{\rm
Re}\,\left<\rho_\varepsilon\right>\,>0$, and second
relationship, clear, take place for real $\rho_\varepsilon$.
Under fulfilment of condition (4.20) due to Theorem 2 the
eigenvalue is of the form
$$
\lambda_\varepsilon= -4\exp\left\{-\frac{4\pi \left(1+\varepsilon
\left<\rho_\varepsilon\Delta^{-1}
\chi\right>\right)}{\varepsilon|Q|\left<\rho_\varepsilon\right>}-2{\bf
C}+O(\varepsilon^\infty) \right\}.\eqno(4.21)
$$

By analogy, due to Theorem 2 the sufficient condition of absence
small eigenvalue  is an existence of $\alpha>0$ such that the
inequalities  (2.8) or (2.9) hold true, those due to (4.19) are
of the form
$$
\hbox{\rm
Re}\,\frac{1+\varepsilon\left<\rho_\varepsilon\Delta^{-1}\chi\right>}
{\varepsilon\left<\rho_\varepsilon\right>}
<-\alpha\varepsilon^{\alpha}+\frac{|Q|\left(\ln 2-{\bf
C}\right)}{2\pi}
$$
or
$$
 \left|\hbox{\rm Im}\,\frac{1+\varepsilon\left<\rho_\varepsilon
\Delta^{-1}\chi\right>} {\varepsilon\left<\rho_\varepsilon\right>}
\right| >\frac{|Q|}{4}+\alpha\varepsilon^{\alpha}.
$$
If quantities $\left<\rho_\varepsilon \Delta^{-1}\chi\right>$
and $\left<\rho_\varepsilon\right>$ are bounded by module as
$\varepsilon\to0$ and $\hbox{\rm
Re}\,\left<\rho_\varepsilon\right>\,<0$, then first of these
inequalities holds a fortiori, and second inequality, obviously,
take place for $\left|\hbox{\rm
Im}\,\left<\rho_\varepsilon\right>\right|>c>0$.

{\bf Remark 3.} Defining concretely $\rho_\varepsilon$ (like it
was done in Examples 1 and 2), by simple relationships (4.20)
and (4.21) it is easy to get more explicit formulas for
eigenvalues (like in Examples 1, 2). In particular, from (4.20)
and (4.21) it follows that if $\rho_\varepsilon\equiv1$ in $Q$,
then
$$
\aligned \lambda_\varepsilon&=
-\varkappa_\varepsilon^2\exp\left\{-\frac{4\pi
}{\varepsilon|Q|^2}
\right\},\\\varkappa_\varepsilon&=2\exp\left\{-\frac{2\pi
\left<\Delta^{-1} \chi\right>}{|Q|^2}-{\bf
C}\right\}+O(\varepsilon^\infty). \endaligned
$$
On the other hand, for the case
$\mathcal{L}_\varepsilon=V_\varepsilon$ considered in Examples 1
and 2 under fulfilment of inequalities (2.7) from Theorem 2 and
(2.6) we deduce the following ''general'' analog of formula
(4.21) for eigenvalues:
$$
\lambda_\varepsilon=
-4\exp\left\{-\frac{4\pi}{\varepsilon\sum\limits_{j=0}^\infty
\left(-\varepsilon\right)^j\left<\left(V_\varepsilon\circ\Delta^{-1}
\right)^jV_\varepsilon\right>}-2{\bf C}+O(\varepsilon^\infty)
\right\},
$$
that is, however, is not such constructive like (4.21) and
formulas obtained in Examples 1 and 2.

{\bf Example 4.} Let
$$
\mathcal{L}_\varepsilon g=\sum\limits_{i,j=1}^2
\frac{\partial}{\partial x_i}\left(a_{ij}\frac{\partial}{\partial
x_j}g\right)+\sum\limits_{i=1}^2\frac{\partial}{\partial
x_i}\left(a_{i}g\right)+\mathcal{L}^\varepsilon g, \eqno(4.22)
$$
where $a_{ij},\,a_i,\, \in C^\infty_0(Q)$,
$$
\mathcal{L}^\varepsilon\overset{def}{=}V_\varepsilon\quad\hbox{\rm
or}\quad
\mathcal{L}^\varepsilon[g]\overset{def}{=}\chi(Q)\left<\rho_\varepsilon
g\right>.
$$
Since $\mathcal{L}_\varepsilon[1]=\mathcal{L}^\varepsilon[1]$
and $\left<\mathcal{L}_\varepsilon
g\right>=\left<\mathcal{L}^\varepsilon g\right>$, by (2.6) the
function $\widetilde{M}(\varepsilon)$, corresponding to the
operator defined by (4.22), coincides with the function
$\widetilde{M}(\varepsilon)$, corresponding to the operator
$\mathcal{L}_\varepsilon=\mathcal{L}^\varepsilon$. Therefore,
all the results established in Examples 1--3 (for operator
$\mathcal{L}_\varepsilon=V_\varepsilon$ and operator defined by
equality
$\mathcal{L}^\varepsilon[g]=\chi(Q)\left<\rho_\varepsilon
g\right>$), are carried over with no changes to the case of
operator defined by equality  (4.22).

{\bf Acknowledgments.} The work has done under support of RFBR
grants (02-01-00693, 00-15-96038) and Ministry of Education of
Russia (E00-1.0-53).

\medskip

\centerline{\bf References}

\noindent [1] {\it L.~D.~Landau, E.~M.~Lifshits.}   Theoretical
physics. Vol.~3. Quantum mechanics. Non-relativistic theory. M.:
Nauka, 1974. (in Russian)

\noindent [2] {\it B.~Simon.} Ann. Phys. 1976. V.~97.
P.~279--288.

\noindent [3] {\it M.~Klaus.} Ann. Phys. 1977. V.~108.
P.~288--300.

\noindent [4] {\it R.~Blankenbecler, M.L.~Goldberger, B.~Simon.}
Ann. Phys. 1977. V.~108. P.~69--78.

\noindent [5] {\it M.~Klaus, B.~Simon.} Ann. Phys. 1980. V.~130.
P.~251--281.

\noindent [6] {\it F.~Bentosela, R.M.~Cavalcanti, P.Exner,
V.A.~Zagrebanov.} J. Phys. A. 1999. V.~32. P.~3029--3039.

\noindent [7] {\it D.E.~Pelinovsky, C.~Sulem.} Theor. Math.
Phys. 2000. V.~122. P.~98-106.

\noindent [8] {\it D.E.~Pelinovsky, C.~Sulem.} Commun. Math.
Phys. 2000. V.~208. P.~713--760.

\noindent [9] {\it P.~Zhevandrov, A.~Merzon.} Trans. AMS
(accepted).

\noindent [10] {\it P.~Zhevandrov, A.~Merzon.} Proc. 3-rd Intern.
ISAAC  Congress. Berlin. 2001 (accepted).

\noindent [11] {\it R.~R.~Gadyl'shin.} Theor. Math. Phys. 2002.
V.~132. P.~976--982.

\noindent [12] {\it I.~S.~Gradshteyn, I.~M.~Ryzhik.} Table of
integrals, series, and products. New York - London - Toronto:
Academic Press, 1980.

\noindent [13] {\it H.~Bateman, A.~Erd\'elyi.}  Higher
transcendental functions. Vol. 2. M.: ''Nauka'', 1974.

\noindent [14] {\it D.~Colton, R.~Kress.} Integral equation
methods in scattering theory. Pure and Applied Mathematics. A
Wiley-Interscience Publication. New York etc.: John Wiley, 1983.

\end{document}